\documentclass[12pt]{article}

\usepackage{amsmath}
\usepackage{amssymb}
\usepackage[margin=3cm]{geometry}
\usepackage{graphicx}
\usepackage{cite}
\numberwithin{equation}{section}
\allowdisplaybreaks
\usepackage{hyperref}

\def\be{\begin{equation}}
\def\ee{\end{equation}}
\def\ba{\begin{eqnarray}}
\def\ea{\end{eqnarray}}

\begin{document}


\title{Asymptotic symmetries in Bondi gauge and the sub-subleading soft graviton theorem}
\author{Bart Horn\thanks{bhorn01@manhattan.edu}}
\date{\normalsize\textit{Department of Physics and Astronomy, Kakos School of Science, \\ Manhattan College, New York, NY 10471, USA}}

\maketitle

\abstract{We investigate asymptotic symmetries which preserve the Bondi gauge conditions but do not preserve the asymptotic falloff conditions for the metric near the null boundary, and their connection to soft graviton theorems for scattering amplitudes.  These include generalized superrotation symmetries parameterized by a smooth vector field $Y^{A}$ obeying $D_{A}Y^{A} = 0$, for which we show that the associated conserved charge can be derived by applying the Noether procedure to the Einstein-Katz action.  We also discuss the connection between asymptotic symmetries and the conserved charge associated with the sub-subleading soft theorem, and we find that in Bondi gauge this charge is generated by the combination of a diffeomorphism together with an extra transformation of the metric.}

\section{Introduction}

Asymptotic symmetries of gauge and gravitational theories, which are gauge (or diffeomorphism) symmetries that do not fall off at infinity, have seen a surge of interest in recent years.  Such symmetries preserve the gauge-fixed action but not the wavefunction, and therefore promote the gauge/diffeomorphism symmetry to a physical one.  For nearly flat spacetimes, asymptotic Killing vectors enlarge the Poincar\'e algebra of symmetries to the infinite-dimensional Bondi-Metzger-Sachs algebra, which may have interesting consequences for the study of flat space holography and the black hole information paradox (see e.g. \cite{Strominger:2017zoo} for a recent review).  Recently in \cite{Strominger:2013jfa, He:2014laa, Kapec:2014opa} it was shown that the Ward identities of BMS symmetries are related to soft graviton theorems for scattering amplitudes, which relate N-pt scattering amplitudes to (N+1)-pt scattering amplitudes in the ``soft'' limit where one of the graviton momenta becomes vanishingly small:
\begin{equation}
\lim_{q \to 0^{+}} \epsilon_{\mu}\epsilon_{\nu}\mathcal{M}^{\mu \nu}(q ; p_1, \cdots, p_N) = \Bigg[ S_{0}(q) + S_{1}(q) + S_2(q) + \cdots \Bigg]\mathcal{M}(p_1, \cdots p_N)\,.
\end{equation}
Here $\mathcal{M}^{\mu \nu}$ and $\mathcal{M}$ refer to scattering amplitudes in momentum space with and without a soft graviton, respectively, and we have used the fact that for a graviton of definite helicity, the graviton polarization tensor can be factorized as $\epsilon_{\mu \nu} = \epsilon_{\mu}\epsilon_{\nu}$.  The leading soft factor is given by Weinberg's soft graviton theorem \cite{Weinberg:1965nx}, and the subleading and sub-subleading soft factors are also given at tree level by
\begin{eqnarray}
\begin{split}
S_{0}(q) &= \frac{\kappa}{2}\sum_{k} \frac{(p_k \cdot \epsilon)^2}{(p_k \cdot q)} \\
S_{1}(q) &= \frac{\kappa}{2}\sum_{k} \frac{(p_k \cdot \epsilon)(\epsilon_{\mu}q_{\nu}J_{k}^{\mu \nu})}{(p_k \cdot q)} \\
S_{2}(q) &= \frac{\kappa}{2}\sum_{k} \frac{(\epsilon_{\mu}q_{\nu}J_{k}^{\mu \nu})(\epsilon_{\rho}q_{\sigma}J_{k}^{\rho \sigma})}{2(p_k \cdot q)}
\end{split}
\end{eqnarray}
where $\kappa^2 = 32 \pi G_N$, $\epsilon^{\mu}$ is the polarization vector of the soft graviton, $J^{\mu \nu}_k = (p^{\mu}_k \partial^{\nu}_{p_k} - p^{\nu}_k \partial^{\mu}_{p_k} + \Sigma_{k}^{\mu \nu})$ is the angular momentum operator including orbital and spin terms, and the sum is taken over the ``hard'' (finite) momenta $p_k$ of the other N particles in the amplitude.  As discussed in \cite{Bern:2014vva}, the leading and subleading soft factor are gauge invariant because of translation and Lorentz invariance, and the form of the subleading and sub-subleading pieces can be derived at tree level using the Taylor expansion and gauge invariance.  Loop corrections at the subleading level and beyond are addressed in e.g. \cite{Bern:2014oka}.

The connection between asymptotic symmetries and soft theorems has many generalizations, including to gauge fields \cite{He:2014cra}, supersymmetric theories and fermionic soft particles \cite{Dumitrescu:2015fej}, to massive particles and symmetries acting at timelike infinity \cite{Campiglia:2015kxa} and to symmetries at spatial infinity (see e.g.\cite{Campiglia:2017mua}).  Connections between multiple soft graviton theorems and the BMS symmetry algebra were discussed in\cite{Distler:2018rwu}.  The soft theorems were identified with the operators of a CFT structure at null infinity in \cite{Pasterski:2016qvg, Pasterski:2017kqt, Donnay:2018neh, Donnay:2020guq, Pasterski:2021fjn, Donnay:2022sdg}.  The literature on asymptotic symmetries and soft theorems is quite extensive, and while the references given here are intended to be illustrative, we emphasize that they are by no means exhaustive.      Asymptotic symmetries have also been investigated in the context of cosmological correlators in expanding spacetimes, in which case the corresponding Ward identities can be understood in terms of relations involving equal-time correlation functions, rather than $\mathcal{S}$-matrix elements \cite{Hinterbichler:2013dpa, Horn:2014rta, Horn:2015dra}.

In this note we study asymptotic symmetries in Bondi gauge that do not preserve the metric falloff conditions near the null boundary, finding a new class of symmetries that have finite charge in the $r \to \infty$ limit, and we explore the relationship between such transformations and the soft graviton theorems.  In the process we clarify and discuss the role of superrotations parameterized by smooth non-holomorphic diffeomorphisms.  We also identify the asymptotic symmetries associated with the sub-subleading soft theorem, and we find that although the conserved charge can be built from pieces of charges generated by asymptotic diffeomorphism symmetries, generating all the pieces together with the desired powers of $r$ in the large-$r$ limit requires an additional non-diffeomorphism transformation of the metric to enforce the Bondi gauge conditions.  This generalizes to Bondi gauge the results found in \cite{Campiglia:2016jdj, Campiglia:2016efb}, in which a correspondence was found between the sub-subleading soft theorem and an asymptotic diffeomorphism symmetry in de Donder gauge.  Although there are many similarities between the two gauges, the form of the allowed transformations are different, and the radial slicing (and therefore the associated large$-r$ expansion) is different as well.  Asymptotic symmetries of spacetime that do not preserve the asymptotic falloff conditions around the Minkowski spacetime metric near the null boundary have also previously been investigated in the context of generalized superrotations in \cite{Campiglia:2014yka}, and for cosmological correlators in Friedmann-Robertson-Walker spacetimes in \cite{Hinterbichler:2013dpa, Horn:2014rta, Horn:2015dra}.  For previous work on the sub-subleading soft theorem and its associated conserved charges, see also \cite{Conde:2016rom}, in which it was identified in Newman-Unti gauge with a part of the supertranslation charge subleading in the large-$r$ expansion, and also especially \cite{Freidel:2021dfs}, which identifies the conserved charge in Bondi gauge and investigates the associated transformation on the phase space of the asymptotic Einstein equations.  In the present work we derive the charge from a complementary perspective. 

This paper is organized as follows: in \S 2 we review the formalism of metric perturbations in Bondi gauge, and we discuss asymptotic symmetries in Bondi gauge and their associated charges.  In \S 3 we review the conserved charge associated with the sub-subleading soft-graviton theorem and show that it commutes with the $\mathcal{S}-$matrix, and in \S 4 we show how it can be generated by a combination of an asymptotic diffeomorphism and an extra metric transformation.  We conclude and indicate further directions in \S 5.

\section{Asymptotic symmetries and charges in Bondi gauge}

The study of asymptotic symmetries for asymptotically Minkowski spacetimes was begun by Bondi, Metzger, van der Burg and Sachs in \cite{Bondi:1962px, Sachs:1962wk, Sachs:1962zza}.  Consider the metric for flat Minkowski spacetime in retarded time coordinates:
\begin{equation}
ds^2 = \eta_{\mu \nu}dx^{\mu} dx^{\nu} = -du^2 - 2du dr + 2r^2 \gamma_{z \bar{z}}dz d\bar{z}
\end{equation}
Here $u = t - r$ is the retarded time coordinate, and $A = \left\{z, \bar{z}\right\}$ are coordinates on the celestial 2-sphere, which has the round metric 
\begin{equation}
\gamma_{z\bar{z}} =  \frac{2}{(1+z \bar{z})^2}\,.
\end{equation}  
These coordinates are ideal for studying the structure at asymptotic null infinity $\mathcal{I}^{+}$ in the limit $r \to \infty$, and a similar description using advanced times $v = t + r$ can be applied to the study of the structure at past null infinity $\mathcal{I}^{-}$ as well.  Allowing fluctuations around this metric, the Bondi gauge conditions are defined by
\begin{equation}
g_{rr} = g_{rA} = 0\,, \qquad \det g_{AB} = r^4 \gamma_{z\bar{z}}^2\,.
\end{equation}

Diffeomorphisms $x^{\mu} \to x^{\mu} + \xi^{\mu}$ cause a shift $\delta g_{\mu \nu} = \nabla_{\left(\mu\right.}\xi_{\left.\nu\right)}$ of the metric, so the diffeomorphisms that preserve Bondi gauge can be shown to be of the form
\begin{equation}
\xi^{\mu} = \Bigg\{ f, -\frac{r}{2}D_{A}Y^{A} + \frac{1}{2}D_{A}D^{A}f, Y^{A}- \frac{1}{r}D^{A}f\Bigg\}\,,
\end{equation}
parameterized by functions $f(u,z,\bar{z}), Y^{A}(u,z,\bar{z})$, which are independent of $r$.  Starting from Minkowski space, these diffeomorphisms generate the metric transformations
\begin{equation}\label{metricperturbations}
\begin{split}
\delta g_{uu} &= \partial_{u}(r D_{A}Y^{A} - 2 f - 2 \gamma^{z\bar{z}}\partial_{z}\partial_{\bar{z}}f)\,, \\ 
\delta g_{ur} &= \left(\frac{1}{2}D_{A}Y^{A} - \partial_{u}f \right)\,, \\
\delta g_{uz} &= r^2 \gamma_{z \bar{z}}\partial_{u}Y^{\bar{z}} + r \partial_{z}\left(\frac{1}{2}D_{A}Y^{A} - \partial_{u} f \right) - \gamma^{z \bar{z}}\partial_{\bar{z}}D_{z}^2 f \,, \qquad \delta g_{u\bar{z}} = c.c.\\
\delta g_{zz} &= 2r^2 \gamma_{z\bar{z}}\partial_{z}Y^{\bar{z}} - 2 r D_{z}^{2}f \,, \qquad \delta g_{\bar{z}\bar{z}} = c.c.\,, \\
\end{split}
\end{equation}
and so further conditions are usually defined in order to preserve the asymptotic form of the metric.  These metric perturbations will also give corrections to the form of the asymptotic Killing vector.  

\subsection{BMS transformations and asymptotically Minkowski spacetimes}

Restricting to the functions
\begin{equation}
f(u, z, \bar{z}) = T(z, \bar{z}) + \frac{u}{2} D_{A}Y^{A} \,, \qquad Y^{z}(u, z, \bar{z}) = Y^{z}(z)\,, \qquad Y^{\bar{z}}(u, z, \bar{z}) = Y^{\bar{z}}(\bar{z})\,,
\end{equation}
where $T(z, \bar{z})$ is the called the supertranslation and $Y^{A}$ is the superrotation, at leading order these transformations generate the metric transformations
\begin{equation}
\begin{split}
\delta g_{uz} &= - \gamma^{z \bar{z}}\partial_{\bar{z}}D_{z}^2 f \,, \qquad \delta g_{u\bar{z}} = c.c.\\
\delta g_{zz} &= - 2 r D_{z}^{2}f \,, \qquad \delta g_{\bar{z}\bar{z}} = c.c. \\
\end{split}
\end{equation}
around Minkowski space.  More specifically, using the Einstein equations, it can be shown that these transformations preserve the form of the asymptotically Minkowski metric
\begin{equation}\label{BMSperturbations}
ds^2 = -e^{2\beta}\left(\left(1-\frac{2m}{r}\right)du^2 + 2dudr\right) -2U_{A}dx^{A}du + g_{AB}dx^{A}dx^{B}\,,
\end{equation}
where to $\mathcal{O}(1/r^2)$, the corrections have the form
\begin{equation}
\begin{split}
e^{2\beta} &= 1 - \frac{1}{16r^2} C_{AB}C^{AB} + \cdots\\
U_{A} &= -\frac{1}{2}D^{B}C_{AB} - \frac{2}{3r}\left(\frac{1}{4}C_{AB}D_{C}C^{BC} + N_{A}\right) + \frac{W_{A}}{r^2} + \cdots\\
g_{AB} &= rC_{AB} +r^2 \gamma_{AB} + \frac{1}{4}C_{CD}C^{CD}\gamma_{AB} + \frac{D_{AB}}{r}\cdots\,,
\end{split}
\end{equation}
and the Bondi gauge conditions fix the form of the asymptotic Killing vector to be
\begin{equation}\label{asympsymm}
\begin{split}
\xi^{u} = f\,, \qquad \xi^{A} = Y^{A} - \frac{1}{2r}D^{A}f + \frac{1}{2r^2}C^{AB}D_{B}f\,,\\
\xi^{r} = -\frac{1}{2}D_{A}Y^{A} + \gamma^{z\bar{z}}\partial_{z}\partial_{\bar{z}}f - \frac{1}{4r}C^{AB}D_{A}D_{B}f + \frac{1}{r}U^{A}D_{A}f\,,
\end{split}
\end{equation}
up to terms of .higher order in $1/r$ and in the metric perturbations.  The subleading metric perturbations contain the quantities $m_B$ and $N_{A}$, which are known respectively as the Bondi mass and the Bondi angular momentum, and they are related to the mass and angular momentum of isolated systems in spacetime.  The Bondi news tensor $N_{AB} = \partial_{u} C_{AB}$ contains information about the outgoing gravitational radiation.  Under a supertranslation or superrotation, the Bondi news tensor receives a nonlinear shift by $\delta N_{zz} = -2\partial_{u}D_{z}^{2} f, \delta N_{\bar{z}\bar{z}} = -2 \partial_{u}D_{\bar{z}}^{2}f$, which can be understood as coming from the addition of soft gravitons, and  a linear (i.e. proportional to $N_{AB}$ itself) shift, which can be shown to be related to a coordinate transformation acting on the hard gravitons.  It was shown in \cite{Strominger:2013jfa, He:2014laa} that the leading part of Weinberg's soft graviton theorem implies that the Ward identities for the supertranslations vanish, and in \cite{Kapec:2014opa} that the Ward identities of the superrotations correspond to the subleading terms in the soft graviton theorem.  

We can find the conserved charge associated with the metric transformations by applying the Noether procedure to the Einstein-Katz action \cite{Katz:1985}
\begin{equation}\label{Katzaction}
S = \frac{1}{16\pi G}\int d^{4}x\, (\sqrt{-g} R - \sqrt{-\bar{g}}\bar{R} + \partial_{\mu}k^{\mu})\,,
\end{equation}
where $\bar{g}_{\mu \nu} = \eta_{\mu \nu}$ is the unperturbed Minkowski metric, and we are including the improvement term on the boundary
\begin{equation}
k^{\mu} = \frac{1}{\sqrt{-g}}\partial_{\nu}(\sqrt{-g}g^{\nu \mu}) = \sqrt{-g}(g^{\mu \nu}\delta \Gamma^{\rho}_{\nu \rho} - g^{\nu \rho}\delta \Gamma^{\mu}_{\nu \rho})\,,
\end{equation}
where the difference $\delta \Gamma = \Gamma - \bar{\Gamma}$ of Christoffel symbols is itself a tensor.  We find that the Noether current has the form $j^{\mu} = \nabla_{\nu}K^{\nu \mu}$ on-shell, where
\begin{equation}\label{Kexpression}
K^{\mu \nu} = \frac{1}{16\pi G}(\sqrt{-g}\nabla^{\left[\mu\right.}\xi^{\left.\nu \right]} - \sqrt{-\bar{g}}\bar{\nabla}^{\left[\mu\right.}\xi^{\left.\nu \right]} + \sqrt{-g}\xi^{\left[\mu\right.}k^{\left.\nu \right]})\,.
\end{equation}
Both the action and the Noether current are manifestly covariant but not background-independent, since the unperturbed metric $\eta_{\mu \nu}$ appears explicitly, but this is perhaps not surprising, given that the asymptotic charges are defined as perturbations around flat Minkowski space.  The fact that the Noether current has only a boundary contribution is consistent with the observation that there are no local observables in a gravitational theory.  The charge can be found by integrating over the future null boundary $\mathcal{I}^{+}$ in the limit $r \to \infty$:
\begin{equation}
Q = \int_{\mathcal{I}^{+}} \ast J = \int_{\mathcal{I}^{+}_{\pm}} \ast K = \int_{\mathcal{I}^{+}_{\pm}} K^{ru}\,,
\end{equation}
where $\mathcal{I}^{+}_{\pm}$ are the slices of the boundary in the limits where $u \to \pm \infty$, and the part finite in the limit $\lim_{r \to \infty}$ is given by\footnote{Note that the coefficient of the term quadratic in $C_{AB}$ differs from the one reported in \cite{Kapec:2014opa}; however, it can be shown that for superrotations this term does not contribute to the action of the hard part on the charge on the graviton operators.  Different conventions for this term are discussed in e.g.\ \cite{Compere:2019gft}.}
\begin{equation}\label{finitecharge}
Q = -\frac{1}{16 \pi G_N}\int d^2 z \gamma_{z\bar{z}}\,\left( 4mf + 2 N_A Y^{A} + \frac{3}{16}Y^{A}D_{A}(C_{CD}C^{CD})\right)\,.
\end{equation}
It can be shown (see \cite{He:2014laa, Kapec:2014opa}, or see e.g. \cite{Distler:2018rwu} for a review) that for ordinary supertranslations and superrotations, the charge commutes with the $\mathcal{S}-$matrix.  Using the Einstein equations, the metric perturbations can be expressed in terms of the Bondi news $N_{AB}$ and its derivatives.  Terms that are linear in the Bondi news correspond to the soft graviton insertion, and terms quadratic and higher generate transformations on the hard modes.  If there is a matter part of the action as well, the Einstein equations also include terms in the charge coming from the matter Noether current $j^{\mu} = T^{\mu \nu}\xi_{\nu}$, contributing the following terms to the charge: 
\begin{equation}\label{finitemattercharge}
Q_{matter} = \int du d^2 z \, \gamma_{z\bar{z}}\, (f T^{(-2)}_{uu} + Y^{A}T^{(-2)}_{uA})\,,
\end{equation} 
where the stress tensor has been expanded as $T_{\mu \nu} = \frac{T^{(-2)}_{\mu \nu}}{r^2} + \frac{T^{(-3)}_{\mu \nu}}{r^3}+ \cdots$.  The matter part of the Noether current can also be derived by working in the approximation of a fixed (non-dynamical) background metric, and applying the Noether procedure to the matter action only.

\subsection{Asymptotic symmetries with finite charges}

The conditions $\partial_{\bar{z}}Y^{z} = \partial_{z}Y^{\bar{z}} = 0$ for superrotations are important for preserving the asymptotic falloff conditions for the metric perturbations in the $r \to \infty$ limit, but they are not needed for satisfying the Bondi gauge conditions.  In \cite{Campiglia:2014yka} the superrotation $Y^{A}(z, \bar{z})$ was promoted to a smooth vector field: this helps simplify the proof of equivalence between the soft theorem and the Ward identity, at the cost of considering transformations that do not preserve the asymptotic Minkowski form of the metric.  In what follows we will start with the asymptotically Minkowski form of the metric given in Eq. \eqref{BMSperturbations} and consider transformations that preserve the Bondi gauge conditions, but not necessarily the asymptotic falloff conditions on the metric perturbations.  Additional metric perturbations will be generated by the transformation, but these are initially set to zero.

Taking the ansatz $f(u,z,\bar{z}) = \int du \frac{1}{2}D_{A}Y^{A}$, the metric transformations generated at leading order around Minkowski space \eqref{metricperturbations} reduce to 
\begin{equation}
\begin{split}
\delta g_{uu} &= O(r) \\ 
\delta g_{uz} \,, \delta g_{u \bar{z}} &=  O(r^2) \\ 
\delta g_{zz}\,, \delta g_{\bar{z}\bar{z}} &=  O(r^2) 
\end{split}
\end{equation}
and further restricting to the subset of transformations of the form $f = 0, D_{A}Y^{A} = 0$ for otherwise arbitrary $Y^{A}(u,z,\bar{z})$, the metric transformations generated reduce to
\begin{equation}
\begin{split}
\delta g_{uz} &= r^2 \gamma_{z \bar{z}}\partial_{u}Y^{\bar{z}} + \cdots \,, \qquad \delta g_{u\bar{z}} = c.c.\\
\delta g_{zz} &= 2r^2 \gamma_{z\bar{z}}\partial_{z}Y^{\bar{z}} + \cdots \,, \qquad \delta g_{\bar{z}\bar{z}} = c.c. \\
\end{split}
\end{equation}
plus terms of higher order in the metric perturbations around Minkowski space.  Starting with the metric in \eqref{BMSperturbations} and applying the Noether procedure to the Einstein-Katz action, we can find the charge associated with this transformation:
\begin{equation}
Q = -\frac{1}{16\pi G_N}\lim_{r \to 0} \int d^{2}z \gamma_{z\bar{z}}(2r U_{A}Y^{A} + 2N_{A}Y^{A})\,.
\end{equation}
For the case where $D_{A}Y^{A} = 0$, we can use the Einstein equations $U_{A} = -\frac{1}{2}D^{B}C_{AB}$, together with the boundary condition $C_{AB} = D_{A}D_{B}C$ adopted in \cite{He:2014laa}, to show that the divergent term will vanish by integration by parts in $z$ and $\bar{z}$. 

The boundary condition can be understood by writing the contribution of $C_{AB}$ to the integral as 
\begin{equation}
C_{zz} = \int du N_{zz} = -\frac{\kappa}{8\pi}\gamma_{z\bar{z}}\lim_{\omega \to 0}\omega \left[a_{+}^{out}(\omega \hat{x}) + a_{-}^{out}(\omega \hat{x})^{\dagger}\right]\,,
\end{equation}
and the same up to a complex conjugate for $C_{\bar{z}\bar{z}}$.  Here we have performed the mode expansion and used the method of steepest descent when performing the integral over $u$ (see e.g. \cite{He:2014laa, Distler:2018rwu}), assuming that $Y^{A}(u,z,\bar{z}) = Y^{A}(z, \bar{z})$ is independent of $u$. It is then straightforward to show using the leading soft-graviton theorem that when this is inserted into the scattering amplitude, the insertion of $C_{zz}$ contributes a factor
\begin{equation}
\langle out | \left[C_{zz}, \mathcal{S}\right] | in \rangle = -\frac{\kappa^2}{8 \pi}\gamma_{z\bar{z}}\sum_{n}\frac{(p_n \cdot \epsilon_{+})^2}{(p_n \cdot q)}\,.
\end{equation}
We can check that this term commutes Using the holomorphic coordinates
\begin{equation}
q^{\mu} = E\left(1, \frac{(z+\bar{z})}{(1+z \bar{z})},\frac{-i(z-\bar{z})}{(1+z \bar{z})},\frac{(1-z\bar{z})}{(1+z \bar{z})}\right)\,, \qquad \epsilon_{+}^{\mu} = \frac{1}{\sqrt{2}}(\bar{z}, 1, -i, -\bar{z}) = \bar{\epsilon}_{-}^{\mu}\,,
\end{equation}
we can show that this is equal to
\begin{equation}
-\frac{\kappa^2}{8 \pi}\gamma_{z\bar{z}}\sum_{n}\frac{(p_n \cdot \epsilon_{+})^2}{(p_n \cdot q)} = -\frac{\kappa^2}{4 \pi}D_{z}^{2}\sum_{n}\Bigg((p_n \cdot q) \ln (p_n \cdot q) - (p\cdot q) \Bigg)\,,
\end{equation}
up to terms that vanish by momentum conservation. Therefore the boundary data obeys $C_{AB} = D_{A}D_{B}C$, and divergent part of the charge vanishes when $\partial_{u}Y^{A} = D_{A}Y^{A} = 0$.  

For terms that depend on a higher power of $u$, such as $Y^{A}(u, z, \bar{z}) = u^{k}\tilde{Y}^{A}(z, \bar{z})$ with $k > 1$, the term containing the soft graviton insertion contributes a factor
\begin{equation}
u^{k}C_{zz} = \int du \partial_{u}(u^{k}C_{zz}) = -\frac{i\kappa}{8 \pi}\gamma_{z\bar{z}}(-i \omega)(i\partial_{\omega})^{k}\left[a_{+}^{out}(\omega \hat{x}) + (-1)^{k}a_{-}^{out}(\omega \hat{x})^{\dagger}\right]\,,
\end{equation}
and the $\partial_{\omega}$ derivatives make the contributions from the subleading soft theorem (as well as the parts of higher order in the soft momentum $\omega$) vanish.  There will still be a contribution from the leading order soft theorem which is divergent as $\omega \to 0$, but as before, this will cancel after integration by parts in $z$ and $\bar{z}$.

It should be emphasized that we have restricted the metric perturbations to the asymptotically Minkowski form in Eq. \eqref{BMSperturbations}, in order to keep the charge finite in the $r \to \infty$ limit.  The finite charge generates a transformation of the metric that will include additional types of metric perturbations.  Turning on all of these metric perturbations before making the transformation, will give additional terms in the Einstein equations and to the Noether charge, some of which are divergent, however, they can be initially set to zero\footnote{We could even set $C_{AB}$ to zero, to simplify the charge further; however, this restricts us to a spacetime without gravitons at the asymptotic boundary.}.

It can be shown using the subleading soft-graviton theorem that the finite part of the charge (Eqs. \ref{finitecharge}) commutes with the $S-$matrix as in \cite{Kapec:2014opa} (also reviewed in \cite{Distler:2018rwu}). Therefore for $D_{A}Y^{A} = 0$, although the asymptotic symmetry transformation does not preserve the usual asymptotically Minkowski form of the metric, it nevertheless gives rise to a finite conserved charge using the same Noether procedure as for the holomorphic superrotation.

\subsection{Summary}



To summarize this section, there exist asymptotic symmetries of Bondi gauge of the form given in Eq. \eqref{asympsymm}, parameterized by the functions $f(u,z,\bar{z})$ and $Y^{A}(u,z,
\bar{z})$, and for $f = 0, D_{A}Y^{A} = 0$, the associated conserved charge derived from the Einstein-Katz action in the metric \eqref{BMSperturbations} is finite on the null boundary as $r \to \infty$ and is given by  
\begin{equation}\label{finitecharge}
Q = -\frac{1}{16 \pi G_N}\int d^2 z \gamma_{z\bar{z}}\,\left( 2 N_A Y^{A}\right)\,.
\end{equation}
This includes generalized superrotation symmetries with $Y^{A} = Y^{A}(z, \bar{z}), D_{A}Y^{A} = 0$, as well as symmetries with nontrivial dependence on $u$.  The existence of additional symmetries of Bondi gauge, and the existence of finite charges in the case where $f = D_{A}Y^{A} = 0$, raises the possibility that these could give rise to additional asymptotic symmetries and corresponding Ward identities.  Nevertheless, since they will not preserve the form of the asymptotically Minkowski metric, they will in general generate transitions between asymptotically Minkowski and more general spacetimes, and therefore it is not necessarily guaranteed that the corresponding Ward identities should have an expression in terms of Minkowski space $\mathcal{S}$-matrix elements.  

We will show, nevertheless, in the sections that follow, that there is a conserved charge associated with the sub-subleading soft graviton theorem for $\mathcal{S}$-matrix elements at tree level (see also \cite{Freidel:2021dfs} for previous work on this topic), and that it is associated with a transformation with $Y^{A}$ linear in $u$ obeying $D_{A}Y^{A} = 0$, together with a part of the generalized superrotation charge that is subleading in the large $r$ limit.  While this is similar to the story in de Donder gauge \cite{Campiglia:2016efb}, in de Donder gauge a pure diffeomorphism is sufficient to generate the charge associated with the sub-subleading charge, while in Bondi gauge that the symmetry requires an extra transformation of the metric in addition to the one generated by the diffeomorphism.

\section{Sub-subleading soft theorem and charge}

In this section we show that at tree level the sub-subleading soft graviton theorem is associated with the conservation of the following charge
\begin{equation}\label{subsubcharge}
\begin{split}
Q^{(2)} = -\frac{1}{16 \pi G_N} \int_{\mathcal{I}^{+}} d^{2}z \gamma_{z\bar{z}} \Bigg[m u^2 D_{A}D_{B}X^{AB} + 2u N_{A}D_{B}X^{AB} \\+ \frac{3u}{16} D_{B}X^{AB}D_{A}(C_{CD}C^{CD})
+ 6X^{AB}D_{AB}\Bigg]
\end{split}
\end{equation}
See \cite{Freidel:2021dfs} for previous work on the charge associated with the sub-subleading soft theorem, from a different perspective and using a different formalism.  In the next section we will discuss the connection between the sub-subleading charge and the asymptotic symmetries investigated in \S 2.

The conservation of the  can be verified by starting with the metric in Eq.\eqref{BMSperturbations}, using the Einstein equations for $m, N_{A}$ and $D_{AB}$, and expanding the result in terms of creation and annihilation operators.  
\begin{equation}
\begin{split}
\partial_{u}m &= \frac{1}{4}D_{A}D_{B}N^{AB} - \frac{1}{8}N_{AB}N^{AB} - 4\pi G_N \lim_{r \to \infty} r^2 T_{uu}\,,\\
\partial_{u}N_A&=\partial_A m-\frac14D_B\left(D^B D^C C_{CA}-D_AD_CC^{BC}\right)\nonumber\\
&+\frac{1}{16}\partial_A\left(N^{BC}C_{BC}\right)-\frac14N^{BC}D_AC_{BC}-\frac14D_B\left(C^{BC}N_{CA}-N^{BC}C_{CA}\right)\\
&-8\pi G_N \lim_{r\to\infty}r^2 T_{uA}\,,\\
\partial_{u} D_{zz} &= \frac{1}{3}D_{z}N_{z} + \frac{1}{2}m C_{zz} + \frac{1}{12}D_{z}C_{zz}D_{z}C^{zz} + \frac{5}{24}C_{zz}D_{z}^{2}C^{zz} - \frac{1}{8}C_{zz}(D^{z})^{2}C_{zz} \\
&\qquad  + \frac{1}{16}D_{z}^{2}(C^{zz}C_{zz}) - 4\pi G_N \lim_{r\to\infty}r^2 T_{zz}\,.
\end{split}
\end{equation}
Here $D_{AB}$ is part of the metric perturbations at sub-subleading order: $g_{AB} = r^{2}\gamma_{AB} + r C_{AB} + \frac{1}{4}C_{CD}C^{CD}\gamma_{AB} + D_{AB}/r + \cdots$, and we note that the Bondi gauge conditions fix $C_{AB}, D_{AB}$ to be traceless.  (Note also that this is different from the notation used in \cite{Barnich:2011mi}, where $D_{AB}$ is used for the $O(r^0)$ term in the metric.  It can be shown using the equations of motion, however, that this piece can be self-consistently set to zero.)

It can be checked that this satisfies $\langle out | \left[Q^{(2)}, \mathcal{S}\right] | in \rangle = 0$ as long as the sub-subleading soft graviton relation is satisfied.  The soft part of the charge can be rewritten as
\begin{equation}
\begin{split}
Q^{(2)}_{S} = -&\frac{1}{16 \pi G_N} \int_{\mathcal{I}^{+}} du d^{2}z \,\gamma_{z\bar{z}} \Bigg[ \frac{1}{4}u^2 D_{A}D_{B}N^{AB} D_{C}D_{D}X^{CD} \\
& \qquad - \frac{1}{4}u^2 N^{C}_{A}\left(D_{C}D_{B}D^{B}D_{D}X^{AD} -D_{C}D_{B}D^{A}D_{D}X^{BD}\right)\Bigg]\\
&= -\frac{1}{16 \pi G_N} \int_{\mathcal{I}^{+}} du d^{2}z \,\gamma_{z\bar{z}} \Bigg[\frac{u^2}{2} D_{z}^2 (D_{z}D_{z}X^{zz})N^{zz} + c.c.\Bigg]
\end{split}
\end{equation}
where in the first line we have integrated by parts and discarded the singular term
\begin{equation}
\Delta Q^{(2)}_{S} = \frac{1}{16 \pi G_N} \int_{\mathcal{I}^{+}} d^{2}z \,\gamma_{z\bar{z}} \frac{\partial}{\partial u}\Bigg[ \frac{1}{4}u^2 C^{C}_{A}\left(D_{C}D_{B}D^{B}D_{D}X^{AD} -D_{C}D_{B}D^{A}D_{D}X^{BD}\right)\Bigg]
\end{equation}
A similar integration by parts had to be performed for the charge associated with the subleading soft theorem (see e.g.\cite{Distler:2018rwu} for details).  We can then use
\begin{equation}
\int du \,u^2 N_{zz} = \frac{\kappa}{8 \pi}\gamma_{z\bar{z}}\lim_{\omega \to 0^{+}}\partial_{\omega}^2 \left(\omega a_{+}(\omega \hat{x}) + \omega a_{-}(\omega \hat{x})^{\dagger}\right)
\end{equation}
to express the Bondi news in terms of soft graviton insertions.  Crossing symmetry will relate the amplitudes involving incoming and outgoing soft gravitons.  Expressing the sub-subleading soft factor in holomorphic coordinates, this is given by
\begin{equation}
\begin{split}
S_{2}(q) &= \frac{\kappa}{2}\sum_{k} \frac{(\epsilon_{\mu}q_{\nu}J_{k}^{\mu \nu})(\epsilon_{\rho}q_{\sigma}J_{k}^{\rho \sigma})}{(p_k \cdot q)} \\
&= -\frac{\kappa}{2}\omega \left(\frac{(\bar{z}-\bar{z}_k)(1+\bar{z}z_k)^2}{(z-z_k)(1+z_k \bar{z}_k)(1+z \bar{z})} E_k \partial_{E_k}^2 + \right. \\
&\qquad \left. + 2 \frac{(\bar{z}-\bar{z}_k)^2 (1+ \bar{z}z_k)}{(z-z_k)(1+z\bar{z})}\left(\partial_{E_k}\partial_{\bar{z}_k}- \frac{1}{E_k}\partial_{\bar{z}_k}\right)\right. \\
&\qquad + \left. \frac{(\bar{z}-\bar{z}_k)^3 (1+z_k \bar{z}_k)}{(z-z_k)(1+z \bar{z})}\frac{1}{E_k}\partial_{\bar{z}_k}^2 + \frac{2(\bar{z}-\bar{z}_k)(1+ \bar{z}z_k)}{(z-z_k)(1+ \bar{z}z)}h_k \partial_{E_k} \right.\\
&\qquad \left. + \frac{2(\bar{z}-\bar{z}_k)^{2}(1+\bar{z}_k z_k)}{(z-z_k)(1+\bar{z}z)} \frac{h_k}{E_k} \partial_{\bar{z}_k} + \frac{(\bar{z}-\bar{z}_k)(1+z_k \bar{z}_k)}{(z-z_k)(1+z\bar{z})}\frac{h_{k}(h_{k} - 1)}{E_k}\right)
\end{split}
\end{equation}
for an outgoing soft graviton of positive helicity.  For an outgoing soft graviton of negative helicity, the expression needs to be complex conjugated, and in the last (spin-squared) term $h_k (h_k -1)$ needs to be replaced by $h_k (h_k + 1)$.  The soft part of the charge therefore gives
\begin{equation}
\begin{split}
\langle out &| \left[Q_{S}^{(2)}, \mathcal{S}\right] | in \rangle = -\frac{1}{16 \pi G_N} \int_{\mathcal{I}^{+}} du d^{2}z \,\gamma_{z\bar{z}} \Bigg[\frac{u^2}{2} D_{z}^2 (D_{z}D_{z}X^{zz})N^{zz} + c.c.\Bigg] \\
&= \frac{\kappa^2}{128 \pi^2 G_N} \int d^{2}z\, \Bigg[ D_{z}^2 (D_{z}D_{z}X^{zz})\Bigg[\frac{(\bar{z}-\bar{z}_k)(1+\bar{z}z_k)^2}{(z-z_k)(1+z_k \bar{z}_k)(1+z \bar{z})} E_k \partial_{E_k}^2 \\ 
 &+ 2 \frac{(\bar{z}-\bar{z}_k)^2 (1+ \bar{z}z_k)}{(z-z_k)(1+z\bar{z})}\left(\partial_{E_k}\partial_{\bar{z}_k}-\frac{1}{E_k}\partial_{\bar{z}_k}\right) + \frac{(\bar{z}-\bar{z}_k)^3 (1+z_k \bar{z}_k)}{(z-z_k)(1+z_k \bar{z}_k)}\frac{1}{E_k}\partial_{\bar{z}_k}^2 \\
 &+ \frac{2(\bar{z}-\bar{z}_k)(1+ \bar{z}z_k)}{(z-z_k)(1+ \bar{z}z)}h_k \partial_{E_k}  + \frac{2(\bar{z}-\bar{z}_k)^{2}(1+\bar{z}_k z_k)}{(z-z_k)(1+\bar{z}z)} \frac{h_k}{E_{k}} \partial_{\bar{z}_k} \\
 &+ \frac{(\bar{z}-\bar{z}_k)(1+z_k \bar{z}_k)}{(z-z_k)(1+z\bar{z})}\frac{h_{k}(h_{k} - 1)}{E_k}\Bigg] + c.c.\Bigg] \langle out | \mathcal{S}| in \rangle
\end{split}
\end{equation}
where the complex conjugate also includes the substitution $h_k (h_k + 1)$ in the last term.  Using $\kappa^2 = 32 \pi G_N$, and integrating by parts and using the Cauchy-Pompieu formula
\begin{equation}
\partial_{z}\left(\frac{1}{(\bar{z}-\bar{z}_k)}\right) = (2\pi)\delta(z-z_k)\,,
\end{equation}
this becomes
\begin{equation}
\begin{split}
\langle out | \left[Q_{S}^{(2)}, \mathcal{S}\right] | in \rangle &= \sum_{k}\left(\frac{1}{4}D_{z_k}^{2}X^{z_k z_k} (E_k \partial^{2}_{E_k} - 2h_k \partial_{E_k} + E_k^{-1} h_k (h_k + 1)) \right.\\ &+ \frac{1}{4}D_{\bar{z}_k}^{2}X^{\bar{z}_k \bar{z}_k} (E_k \partial^2_{E_k} + 2h_k\partial_{E_k} + {E_{k}}^{-1}h_k (h_k - 1)) \\
&- D_{z_k}X^{z_k z_k}\left(\partial_{E_k} - \frac{(1+h_k)}{E_k}\right)(\partial_{z_k} - h_k \Omega_{z_k}) \\& - D_{\bar{z}_k}X^{\bar{z}_k \bar{z}_k}\left(\partial_{E_k} - \frac{(1-h_k)}{E_k}\right)(\partial_{\bar{z}_k} + h_k \Omega_{\bar{z}_k}) \\
+ \frac{3}{2}X^{z_k z_k} & \left.\frac{1}{E_k}(\partial_{z_k} - h_k \Omega_{z_k})^2 + \frac{3}{2}X^{\bar{z}_k \bar{z}_k}\frac{1}{E_k}(\partial_{\bar{z}_k} + h_k \Omega_{\bar{z}_k})^2 \right)\langle out | \mathcal{S}| in \rangle 
\end{split}
\end{equation}
where the derivatives act on the hard amplitude, and $\Omega_{z} = \frac{1}{2}\Gamma^{z}_{zz}$ is the spin connection.  

The quadratic (hard) part of the gravitational charge is given by
\begin{equation}
\begin{split}
Q^{(2)}_{H} = - \frac{1}{16 \pi G_N}\int du d^{2}z \,\gamma_{z\bar{z}}\Bigg[- \frac{u^2}{8}N_{AB}N^{AB}D_{C}D_{D}X^{CD}
+ \frac{u}{8}\partial_{A}(N_{CD}C^{CD})D_{B}X^{AB} \\ - \frac{u}{2}N^{CD}D_{A}C_{CD}D_{B}X^{AB} - \frac{u}{2}D_{C}(C^{CD}N_{DA} - N^{CD}C_{DA})D_{B}X^{AB}
+ 3m C_{AB}X^{AB} \\+ \frac{1}{4}D_{A}C_{CD}D_{B}C^{CD}X^{AB} + \frac{5}{4}C_{AC}D_{B}D_{D}C^{CD}X^{AB}
- \frac{3}{4}C_{AC}D^{C}D^{D}C_{BD}X^{AB} \\+ \frac{3}{16}D_{A}D_{B}u \partial_{u}(C_{CD}C^{CD})X^{AB} \Bigg]\\
=- \frac{1}{16 \pi G_N}\int du d^{2}z \,\gamma_{z\bar{z}}\Bigg[- \frac{u^2}{8}N_{AB}N^{AB}D_{C}D_{D}X^{CD}
- \frac{u}{4}\partial_{A}(N_{CD}C^{CD})D_{B}X^{AB} \\ - \frac{u}{2}N^{CD}D_{A}C_{CD}D_{B}X^{AB} - \frac{u}{2}D_{C}(C^{CD}N_{DA} - N^{CD}C_{DA})D_{B}X^{AB}\\
+ \frac{3}{4}D_{C}D_{D}C^{CD}C_{AB}X^{AB} + \frac{1}{4}D_{A}C_{CD}D_{B}C^{CD}X^{AB} + \frac{5}{4}C_{AC}D_{B}D_{D}C^{CD}X^{AB}
\\- \frac{3}{4}C_{AC}D^{C}D^{D}C_{BD}X^{AB} \Bigg]
\end{split}
\end{equation}
Using the mode expansions
\begin{equation}
\begin{split}
C_{zz} = \frac{i \kappa}{8 \pi^2}\gamma_{z\bar{z}}\int_{0}^{\infty}d\omega \, \left(a_{-}^{\dagger}e^{i \omega u} - a_{+}e^{-i \omega u}\right)\,, \\ N_{zz} = -\frac{\kappa}{8 \pi^2}\gamma_{z\bar{z}}\int_{0}^{\infty}d\omega \, \omega\left(a_{-}^{\dagger}e^{i \omega u} + a_{+}e^{-i \omega u}\right)\,,
\end{split}
\end{equation}
it can be shown that the hard part of the gravitational charge takes on the form
\begin{equation}
\begin{split}
Q^{(2)}_{H} = \frac{1}{16 \pi^3}\int d^{2}z \,\gamma_{z\bar{z}}\int_{0}^{\infty} d\omega \,\Bigg[-\frac{1}{4}D_{A}D_{B}X^{AB}(a_{+}^{\dagger}\omega \partial_{\omega}^{2}a_{+}+a_{-}^{\dagger}\omega \partial_{\omega}^{2}a_{-}) \\
+ D_{B}X^{AB}(a_{+}^{\dagger}(-1 + \omega \partial_{\omega})D_{A}a_{+}+a_{-}^{\dagger}(-1 + \omega \partial_{\omega})D_{A}a_{-})\\
- \frac{3}{2}X^{AB}(a_{+}^{\dagger}D_{A}D_{B}a_{+} + a_{-}^{\dagger}D_{A}D_{B}a_{-})
+ D_{z}X^{zz}(-2 a_{+}^{\dagger}D_{z}a_{+} + 2 a_{-}^{\dagger}D_{z}a_{-}) + c.c. \\ + D_{z}^{2}X^{zz}a_{+}^{\dagger}\left(-\frac{3}{4} + \omega \partial_{\omega}\right)a_{+} + D_{z}^{2}X^{zz}a_{-}^{\dagger}\left(\frac{1}{4} - \omega \partial_{\omega}\right)a_{-} + c.c.\Bigg]
\end{split}
\end{equation}
and that acting on outgoing graviton operators, the commutator
\begin{equation}
\left[Q_{H}^{(2)}, a_{\pm}(E_k \hat{x}_k)\right]
\end{equation}
generates a transformation that is nearly consistent with the soft-graviton theorem.  Only the helicity squared terms disagree, and the missing piece in the hard part of the charge that would be needed to fix this is given by
\begin{equation}
\Delta Q^{(2)}_{H} = \frac{1}{16 \pi^3}\int d^{2}z \,\gamma_{z\bar{z}}\int_{0}^{\infty} d\omega \,\Bigg[-\frac{3}{4}D_{A}D_{B}X^{AB}(a_{+}^{\dagger}a_{+} + a_{-}^{\dagger}a_{-})\Bigg]\,.
\end{equation}Note that the operators $a_{+}, a_{-}$ create outgoing gravitons with helicities $+2$ and $-2$, respectively.)

If there is a matter sector present, the matter part of the hard charge is given by
\begin{equation}
Q^{(2)}_{H, matter} = \lim_{r \to \infty} r^2 \int du d^{2}z \, \gamma_{z\bar{z}}\, \left(\frac{1}{4}u^2 D_{A}D_{B}X^{AB}T_{uu} + u D_{B}X^{AB}T_{uA} + \frac{3}{2}X^{AB} T_{AB}\right)
\end{equation}
For a massless scalar field $T_{\mu \nu} = \partial_{\mu}\phi \partial_{\nu}\phi - \frac{1}{2}(\partial \phi)^2 g_{\mu \nu}$, and we can make the asymptotic expansion $\phi \approx \frac{\varphi(u, z, \bar{z})}{r} + \cdots $.  Using the mode expansion near null infinity and using the stationary phase approximation, we have
\begin{equation}
\phi = 
\frac{i}{8 \pi^2 r}\int_{0}^{\infty} d\omega \, (a^{\dagger}(\omega \hat{x})e^{i \omega u} - a(\omega \hat{x}) e^{-i\omega u})\,,
\end{equation}
and therefore the matter part of the charge becomes
\begin{equation}
\begin{split}
Q^{(2)}_{H, matter} &= \\ \int &du d^{2}z \, \gamma_{z\bar{z}}\, \left(\frac{1}{4}u^2 D_{A}D_{B}X^{AB}\varphi'^2 + u D_{B}X^{AB}\varphi' \partial_{A}\varphi + \frac{3}{2}X^{AB} \partial_{A}\varphi \partial_{B}\varphi\right)\\
&= \frac{1}{16 \pi^3}\int d^{2}z \gamma_{z\bar{z}}\int_{0}^{\infty} d\omega \Bigg[-\frac{1}{4}D_{A}D_{B}X^{AB}a^{\dagger}\omega^2 \partial_{\omega}^{2}a + D_{B}X^{AB}a^{\dagger}\omega \partial_{\omega}D_{A}a\\
&- D_{B}X^{AB}a^{\dagger}D_{A}a - \frac{3}{2}X^{AB}a^{\dagger}D_{A}D_{B}a + \frac{1}{2}D_{A}D_{B}X^{AB} a^{\dagger}a  \Bigg]\,.
\end{split}
\end{equation}
All terms inside the brackets except for the last one generate the expected transformation of creation and annihilation operators to cancel the contribution of the soft part of the charge.  

Therefore, to ensure that the soft and hard contributions to the charge will indeed commute with the $\mathcal{S}-$matrix, we need to restrict to transformations of the form
\begin{equation}
D_{A}D_{B}X^{AB} = 0.
\end{equation}
In this case the problematic terms in the gravitational and matter charges drop out, and $Q^{(2)} = Q^{(2)}_{S} + Q^{(2)}_{H}$ commutes with the $\mathcal{S}-$matrix.

\section{Sub-subleading charge and asymptotic symmetries}

In this section we explore the connection between the conserved charge in Eq. \eqref{subsubcharge} and the asymptotic symmetries of \S 2.  We begin with the metric in Eq. \eqref{BMSperturbations} and consider the following symmetry of Bondi gauge:
\begin{equation}
f(u,z,\bar{z}) = 0\,, \qquad Y^{A}(u, z, \bar{z}) = u D_{B}X^{AB}\,,
\end{equation}
where $D_{A}D_{B}X^{AB} = 0$.  From Eq.\ref{finitecharge}, the associated charge is given by
\begin{equation}\label{Q2partialcharge}
Q = -\frac{1}{16 \pi G_N}\int d^2 z \,\gamma_{z\bar{z}}\left( 2 uN_A D_{B} X^{AB}\right)\,.
\end{equation}
This agrees with a part of the conserved sub-subleading charge described in the previous section, with $D_{A}D_{B}X^{AB} = 0$; however, the terms proportional to $D_{AB}X^{AB}$ are missing.  

Following the example of \cite{Campiglia:2016efb}, we seek to identify the missing terms with terms proportional to a superrotation charge.  We consider therefore also the transformation $f = 0$, $Y^{A} = X^{A}(z, \bar{z})$, where $D_{A}X^{A} = 0$.  The charge is of the form
\begin{equation}
Q^{(1)} = -\frac{1}{16 \pi G_N} \int d^{2}z\,\gamma_{z\bar{z}}\left(2 N_{A} X^{A}\right)\,
\end{equation}
and using the equations of motion, this contains both soft and hard contributions to the gravitational charge, as well as a matter part of the form
\begin{equation}
Q^{(1)}_{matter} = \lim_{r \to \infty}r^2\int d^{2}z\,\gamma_{z\bar{z}}\left(T^{r}_{\,\,A} X^{A}\right)\,.
\end{equation}
The finite part of $Q^{(1)}$ will commute with the $\mathcal{S}-$matrix, as described in \S 2, and we need also the part of the charge subleading in the $r \to \infty$ limit:
\begin{equation}\label{subleadingsuperrotation}
\begin{split}
Q^{(1)}_{subleading} &= \frac{1}{r}\frac{1}{16 \pi G_N}\int d^{2}z \,\gamma_{z\bar{z}} \left(4W_A  X^A + \frac{2}{3}C_{AB}N^{A}X^{B}\right) \\
&= \frac{1}{r}\frac{1}{16 \pi G_N}\int du d^{2}z \gamma_{z\bar{z}}\, \left(4\partial_{u}W_A  X^A + \frac{2}{3}\partial_{u}\left(C_{AB}N^{A}\right)X^{B}\right)
\end{split}
\end{equation}
Here the quantity $W_{A}$ comes from the $g_{uA}$ terms in the metric: $g_{uA} = \frac{1}{2}D^{B}C_{AB} + \frac{2}{3r}N_{A} + \frac{1}{6r}C_{AB}D_{C}C^{BC} - \frac{1}{r^2}W_{A} + \cdots$.  We can rearrange these terms using the Einstein equations
\begin{equation}
\begin{split}
16 \pi &G_N (T^{(-3)}_{uz} - T^{(-3)}_{rz}) = 4\partial_{u} W_{z} + \gamma^{z\bar{z}}\partial_{u}\partial_{\bar{z}}D_{zz} - \frac{2}{3}\gamma^{z\bar{z}}\left(D_{z}\partial_{\bar{z}}N_{z} - D_{z}^{2} N_{\bar{z}}\right) - \frac{4}{3}N_{z} \\
&+ \frac{2}{3}\partial_{u}(C_{zz}N^{z})- 2D^{z}(mC_{zz}) - \frac{1}{3}C_{zz}D_{z}C^{zz} + \frac{1}{2}C_{zz}(D^{z})^{3}C_{zz} - \frac{1}{2}C_{zz}D^{z}D_{z}^{2}C^{zz}\\
&+ \frac{1}{2}((D^{z})^{2}C_{zz}-D_{z}^{2}C^{zz})D^{z}C_{zz} -\frac{1}{6}D_{z}(D^{z}(C_{zz}D_{z}C^{zz}) - D_{z}(C^{zz}D^{z}C_{zz}))\,,\\
8 \pi &G_N T^{(-2)}_{zz} = -2 \partial_{u}D_{zz} + \frac{2}{3}D_{z}N_{z} + m C_{zz} + \frac{1}{6}D_{z}C_{zz}D_{z}C^{zz} + \frac{5}{12}C_{zz}D_{z}^{2}C^{zz} \\
&- \frac{1}{4}C_{zz}(D^{z})^{2}C_{zz} + \frac{1}{8}D_{z}^{2}(C^{zz}C_{zz})\,,
\end{split}
\end{equation}
where $T_{\mu \nu} = \frac{1}{r^2}T^{(-2)}_{\mu \nu} + \frac{1}{r^3}T^{(-3)}_{\mu \nu} + \cdots $ are the terms in the $1/r$ expansion of the stress tensor.  Making frequent use of the condition $D_{A}X^{A} = 0$, the expression in Eq. \eqref{subleadingsuperrotation} can be simplified step by step:
\begin{align}
Q^{(1)}_{subleading}
&= \frac{1}{r}\frac{1}{16 \pi G_N}\int du d^{2}z \,\gamma_{z\bar{z}} \left(4\partial_{u}W_A  X^A + \frac{2}{3}\partial_{u}\left(C_{AB}N^{A}\right)X^{B}\right) \nonumber \\
= \frac{1}{r}&\frac{1}{16 \pi G_N}\int du d^{2}z \,\gamma_{z\bar{z}}\Bigg(16 \pi G_N (T^{(-3)}_{uz}-T^{(-3)}_{rz})X^{z} - \partial_{u}D^{z}D_{zz}X^{z} \nonumber \\ &+ \frac{2}{3}\gamma^{z\bar{z}}\left(D_{z}\partial_{\bar{z}}N_{z} - D_{z}^{2} N_{\bar{z}}\right)X^{z} + \frac{4}{3}N_{z}X^{z}
+ 2D^{z}(mC_{zz})X^{z} \nonumber \\ &- \frac{1}{2}C_{zz}(D^{z})^{3}C_{zz}X^{z} + \frac{1}{2}C_{zz}D^{z}D_{z}^{2}C^{zz}X^{z} - \frac{1}{2}((D^{z})^{2}C_{zz}-D_{z}^{2}C^{zz})D^{z}C_{zz}X^{z} \nonumber \\ &+ \frac{1}{6}D_{z}(D^{z}(C_{zz}D_{z}C^{zz}) - D_{z}(C^{zz}D^{z}C_{zz}))X^{z} + \frac{1}{3}C_{zz}D_{z}C^{zz}X^{z} + c.c.\Bigg) \nonumber \\
= \frac{1}{r}&\frac{1}{16 \pi G_N}\int du d^{2}z \,\gamma_{z\bar{z}}\Bigg(16 \pi G_N (T^{(-3)}_{uz}-T^{(-3)}_{rz})X^{z} - \partial_{u}D^{z}D_{zz}X^{z} + \frac{4}{3}D^{z}(D_{z}N_{z})X^{z} \nonumber \\
&+ 2D^{z}(mC_{zz})X^{z} - \frac{1}{2}D^{z}(C_{zz}(D^{z})^{2}C_{zz})X^{z} + \frac{1}{2}D^{z}(C_{zz}(D_{z}^{2})C^{zz})X^{z} \nonumber \\ &+ \frac{1}{3}D^{z}(D_{z}(C_{zz}D_{z}C^{zz}))X^{z} + c.c.\Bigg) \nonumber \\
= \frac{1}{r}&\frac{1}{16 \pi G_N}\int du d^{2}z \,\gamma_{z\bar{z}}\Bigg(16 \pi G_N (T^{(-3)}_{uz}-T^{(-3)}_{rz})X^{z} + 16 \pi G_N D^{2}T^{(-2)}_{zz}X^{z} \nonumber \\& + 3\partial_{u}D^{z}D_{zz}X^{z} + c.c.\Bigg)
\end{align}
The stress tensor contributions to the charge can be simplified using the conservation law $\nabla^{\mu}T_{\mu \nu} = 0$:
\begin{equation}
\begin{split}
Q^{(1)}_{H, matter, subleading} &= \frac{1}{r}\int d^{2}z \,\gamma_{z\bar{z}}\left((T^{(-3)}_{uz}-T^{(-3)}_{rz})X^{z} + c.c.\right) \\
&= \frac{1}{r}\int d^{2}z \,\gamma_{z\bar{z}}\left(\left(\partial_{u}T^{(-4)}_{rz} - D^{\bar{z}}T^{(-2)}_{\bar{z}z} - D^{z}T^{(-2)}_{zz}\right)X^{z} + c.c\right) \\
&= - \frac{1}{r}\int d^{2}z \,\gamma_{z\bar{z}} \left(D^{z}T^{(-2)}_{zz} \right)X^{z} + c.c.\,,
\end{split}
\end{equation}
where we have integrated by parts and used the condition $D_{A}X^{A} = 0$ to get rid of the $T^{(-2)}_{z\bar{z}}$ term, and discarded the total $u$-derivative.  This is justified if the matter fields vanish at $u \pm \infty$, as we expect it will for localized massless particle insertions.  The expression in Eq. \eqref{subleadingsuperrotation} then becomes
\begin{equation}
\begin{split}
Q^{(1)}_{subleading} &= \frac{1}{r}\frac{1}{16 \pi G_N}\int du d^{2}z \,\gamma_{z\bar{z}} \left(4\partial_{u}W_A  X^A + \frac{2}{3}\partial_{u}(C_{AB}N^{A})X^{B}\right) \\
&= \frac{1}{r}\frac{1}{16 \pi G_N}\int du d^{2}z \,\gamma_{z\bar{z}} \left(3 \partial_{u}D^{z}D_{zz} X^{z} + c.c.\right)\,.
\end{split}
\end{equation}
This can be combined with the charge derived in Eq. \eqref{Q2partialcharge}, with $Y^{A}$ given by $Y^{z} = D_{z}D^{z}X^{z}, Y^{\bar{z}} = D_{\bar{z}}D^{\bar{z}}X^{\bar{z}}$, which obeys $D_{A}Y^{A} = 0$ because $D_{A}X^{A} = 0$.  This charge is given by
\begin{equation}
\begin{split}
Q = -&\frac{1}{16 \pi G_N} \int_{\mathcal{I}^{+}} d^{2}z \,\gamma_{z\bar{z}} \Bigg[u N_{A}\left(\Delta X^{A} + X^{A}\right )\Bigg]\,,\\
\end{split}
\end{equation}
where we have used the identity $D_{z}D^{z}X^{z} = \frac{1}{2}\left(\Delta X^{z} + X^{z}\right)$. 
Adding this to the charge from the non-flat superrotation in the linear combination $Q - 2 r Q^{(1)}$, the divergent part of the charge in the limit $r \to \infty$ is proportional to a generalized superrotation with $D_{A}X^{A} = 0$ ad therefore commutes with the $\mathcal{S}-$matrix, and the finite part of the charge in the limit $r \to \infty$ becomes
\begin{equation}
\begin{split}
(Q^{(2)} - 2r Q^{(1)})_{finite} &= -\frac{1}{16 \pi G_N} \int d^{2}z \gamma_{z\bar{z}}\, \Bigg(2 u N_{A} D_{A}(D^{A}X^{B})_{TF} + 6 D_{AB}(D^{A}X^{B})_{TF}\Bigg)\,,
\end{split}
\end{equation}
where $(D^{A}X^{B})_{TF}$ refers to the trace-free part of this quantity.  We recognize the finite part of the charge as the charge \eqref{subsubcharge} from in the previous section, which was shown to commute with the $\mathcal{S}-$matrix.  

The charge $-2r Q^{(1)}$ is not generated by a pure diffeomorphism, since the transformation $\xi^{A} = -2 r X^{A}$ which would be required generates a shift $\delta g_{rA}$ that takes us out of Bondi gauge.  Nevertheless, performing this transformation while also transforming the metric so as to enforce the constraint $\delta g_{rA} = 0$ will generate a symmetry of the asymptotic equations of motion.  To see this, we can apply the Noether procedure to the Einstein-Katz action with the transformation $\xi^{A} = -2 r_0 X^{A}$, where $r_0$ is a constant, with value equal to $r_0 = r$ for some large but finite value of $r$.  This is a symmetry of Bondi gauge, and the corresponding charge will act as a symmetry upon the asymptotic data (which does not depend on $r$).  There is also a divergent part proportional to $r_0$ times the finite part of $Q^{(1)}$.  This is not a problem, however, since we already know this this commutes with the $\mathcal{S}-$matrix due to the subleading soft theorem.  Setting $r_0 = r$ and taking the $r \to \infty$ limit, we obtain a charge that is not generated by the diffeomorphism alone; however, it will still act on the boundary data in the same way, and therefore it will generate a transformation between solutions of the equations of motion.

To summarize the results of this section, we have shown that a transformation generated by 
\begin{equation}
\begin{split}
\xi^{A} &= -2 r X^{A} + \frac{u}{2}(\Delta + 1)X^{A} + \cdots\,,\\
\delta g_{ur} &= U_{A}X^{A} - \frac{2}{3r}N_{A}X^{A} - \frac{1}{6r}C_{AB}D_{C}C^{BC}X^{A} + \frac{1}{r^2}W_{A}X^{A} + \cdots\\
\delta g_{rA} &= -r^2 \gamma_{AB}X^{B} - r C_{AB}X^{B} - \frac{D_{AB}X^{B}}{r} - \frac{1}{2}C_{AB}C^{BC}X_{C} + \cdots 
\end{split}
\end{equation}
where $D_{A}X^{A} = 0$ and $\Delta = D_{A}D^{A}$ is the Laplacian on the two-sphere, and the metric transformation is in addition to the one already generated by the diffeomorphism, generates the sub-subleading charge.  This charge includes a term which is proportional to $r$  multiplied by a generalized superrotation charge, which therefore does not interfere with the symmetry of the $\mathcal{S}-$matrix.  As stated above, we emphasize that this is not simply a diffeomorphism symmetry, because the constraint $\delta g_{rA} = 0$ is enforced by an additional transformation of the metric.  This can be compared to the situation in de Donder gauge, where the sub-subleading charge is generated by the diffeomorphism 
\begin{equation}
\xi^{A} = r X^{A} + \frac{u}{4}(\Delta + 5)X^{A} + \cdots\,, 
\end{equation}
found in \cite{Campiglia:2016efb}.  This symmetry also generates a divergent piece that is proportional to a generalized superrotation charge, which commutes with the $\mathcal{S}-$matrix, and a finite piece associated with the sub-subleading soft graviton theorem.  We have therefore identified the charge associated with the sub-subleading soft theorem with a combination of charges derived from asymptotic symmetries, extending the analysis of \cite{Campiglia:2016efb} to Bondi gauge.

While it is perhaps not surprising that there are corresponding transformations in both gauges that are associated with the sub-subleading soft theorem, it is interesting that they correspond to pure diffeomorphism symmetries in some gauges but not in others.  It is furthermore not guaranteed that they had to operate at the same order in the $1/r$ expansion, since the radial slicing is different between Bondi and de Donder gauge.  In \cite{Conde:2016rom} it was found that in Newman-Unti gauge the subleading and sub-subleading soft graviton theorems were associated with the part of the supertranslation charge subleading in the $1/r$ expansion.  It would be interesting to understand the symmetry structure from a more gauge-invariant perspective; see e.g. \cite{Hinterbichler:2013dpa, Horn:2014rta} for a discussion of similar issues in the context of cosmological correlators. 

\section{Conclusions and Open Questions}







In this paper, we have investigated asymptotic symmetries of Bondi gauge and their conserved charges, and their connection with the conserved charge associated with the sub-subleading soft graviton theorem at tree level.  This extends the analysis of \cite{Campiglia:2016efb} in de Donder gauge, where the corresponding symmetry is a pure diffeomorphism, to Bondi gauge, where it is not.  Nevertheless, individual pieces of the charge associated with the sub-subleading soft theorem can be associated with those generated by asymptotic diffeomorphism symmetries.  Our results also complement those of \cite{Freidel:2021dfs}, where the sub-subleading charge and symmetry are investigated from the perspective of the asymptotic phase space.  

Since the sub-subleading soft theorem receives loop corrections, we expect that the symmetry transformation will as well, and it would be interesting to study this further, and to understand whether there is a quantum corrected version of the symmetry transformation and of the charge.  Some discussion of the origins of the corrections to the sub-subleading soft charge can be found in \cite{Freidel:2021dfs}.  It would also be interesting to study the contribution the sub-subleading charge $Q^{(2)}$ makes to the BMS algebra and its extensions, extending the analysis of the BMS charge algebra and its extensions begun in e.g. \cite{Distler:2018rwu, Freidel:2021dfs, Donnay:2020guq, Donnay:2021wrk}.

As a corollary of this analysis, we have also made it clear how the superrotation transformation can be extended beyond (anti)holomorphic transformations of the form $Y^{z}(z)$, $Y^{\bar{z}}(\bar{z})$ to include smooth diffeomorphisms $Y^{A}(z, \bar{z})$ that obey $D_{A}Y^{A} = 0$, by using the same (Einstein-Katz) action to derive the conserved charge directly using the Noether procedure.  It was previously noted in \cite{Campiglia:2014yka} (see also \cite{Distler:2018rwu}) that  smooth functions for $Y^{A}$ improve the proof of the equivalence between the soft theorem and the Ward identity, and smooth functions may generalize more readily to the symmetries of the near-horizon limit for black holes as well\cite{Penna:2015gza, Donnay:2016ejv}.

It would be interesting to understand whether and how the analysis here can be extended beyond transformations with $D_{A}Y^{A} = 0$.  Our analysis in \S 3 indicates that some additional terms in the charge will be required.  Moreover, it is less clear how to generate the charge in this case.  One possibility might be to consider additional improvement terms to the Einstein-Katz action.  In \cite{Freidel:2021dfs} it was proposed that the charge associated with the sub-subleading soft theorem could be generated by a ``pseudo-vector'' which generates a symmetry transformation containing terms with more than one derivative, and it would be interesting to explore the origin and nature of such objects further.

The analysis in this paper also makes it clear that there exist additional asymptotic transformations of Bondi gauge with charges that remain finite in the $r \to \infty$ limit. These are parameterized by functions $Y^{A}(u,z,\bar{z})$ which obey $D_{A}Y^{A} = 0$ but which have arbitrary $u-$dependence.  We have attempted to generalize the procedure here to higher order in $u$ (for instance, to transformations of the form $Y^{A} = u^2 X^{A}(z, \bar{z})$ with $D_{A}X^{A} = 0$) and connect these to soft theorems, but so far without success -- the soft theorem beyond sub-subleading order is not fully fixed by gauge invariance even at tree level \cite{Bern:2014vva}, and attempting to replicate the analysis of \S 4 for this choice of $Y^{A}$ generates terms in the charge that cannot be simply removed by a term proportional to previously known charges.  It would be interesting to study further the meaning of these symmetries and the corresponding Ward identities.  Since these do not preserve the form of the asymptotically flat spacetime metric, as the BMS symmetries do, they should perhaps not necessarily be expected to generate symmetry transformations that can be simply described in terms of the $\mathcal{S}-$matrix -- for example, a sub-sub-subleading soft graviton may generate a transformation that relates a flat space amplitude to a correlation function in a spacetime with a different asymptotic structure.
From this perspective, it would be interesting to understand better from this perspective why a description of the Ward identity in terms of flat space $\mathcal{S}-$matrix elements can still work at the level of the sub-subleading soft theorem.   

\section*{Acknowledgments}

It is a pleasure to thank M.~Campiglia, L.~Donnay, R.~Flauger, L.~Hui, A.~Joyce, A.~Nicolis, R.~Penna, and C.~Zukowski for very helpful discussions, and we further thank M.~Campiglia and R.~Penna for detailed comments on a draft manuscript.  BH is also supported by the National Science Foundation under grant PHY-2210475.


\begin{thebibliography}{10}


\bibitem{Strominger:2017zoo}
A.~Strominger,
``Lectures on the Infrared Structure of Gravity and Gauge Theory,''
[arXiv:1703.05448 [hep-th]].

\bibitem{Strominger:2013jfa}
A.~Strominger, 
``On BMS Invariance of Gravitational Scattering,''
JHEP \textbf{07}, 152 (2014)
doi:10.1007/JHEP07(2014)152
[arXiv:1312.2229 [hep-th]].

\bibitem{He:2014laa}
T.~He, V.~Lysov, P.~Mitra and A.~Strominger,
``BMS supertranslations and Weinberg\textquoteright{}s soft graviton theorem,''
JHEP \textbf{05}, 151 (2015)
doi:10.1007/JHEP05(2015)151
[arXiv:1401.7026 [hep-th]].

\bibitem{Kapec:2014opa}
D.~Kapec, V.~Lysov, S.~Pasterski and A.~Strominger,
``Semiclassical Virasoro symmetry of the quantum gravity $ \mathcal{S}$-matrix,''
JHEP \textbf{08}, 058 (2014)
doi:10.1007/JHEP08(2014)058
[arXiv:1406.3312 [hep-th]].

\bibitem{Weinberg:1965nx}
S.~Weinberg,
``Infrared photons and gravitons,''
Phys. Rev. \textbf{140}, B516-B524 (1965)
doi:10.1103/PhysRev.140.B516

\bibitem{Bern:2014vva}
Z.~Bern, S.~Davies, P.~Di Vecchia and J.~Nohle,
``Low-Energy Behavior of Gluons and Gravitons from Gauge Invariance,''
Phys. Rev. D \textbf{90}, no.8, 084035 (2014)
doi:10.1103/PhysRevD.90.084035
[arXiv:1406.6987 [hep-th]].

\bibitem{Bern:2014oka}
Z.~Bern, S.~Davies and J.~Nohle,
``On Loop Corrections to Subleading Soft Behavior of Gluons and Gravitons,''
Phys. Rev. D \textbf{90}, no.8, 085015 (2014)
doi:10.1103/PhysRevD.90.085015
[arXiv:1405.1015 [hep-th]].

\bibitem{He:2014cra}
T.~He, P.~Mitra, A.~P.~Porfyriadis and A.~Strominger,
``New Symmetries of Massless QED,''
JHEP \textbf{10}, 112 (2014)
doi:10.1007/JHEP10(2014)112
[arXiv:1407.3789 [hep-th]].

\bibitem{Dumitrescu:2015fej}
T.~T.~Dumitrescu, T.~He, P.~Mitra and A.~Strominger,
``Infinite-dimensional fermionic symmetry in supersymmetric gauge theories,''
JHEP \textbf{08}, 051 (2021)
doi:10.1007/JHEP08(2021)051
[arXiv:1511.07429 [hep-th]].

\bibitem{Campiglia:2015kxa}
M.~Campiglia and A.~Laddha,
``Asymptotic symmetries of gravity and soft theorems for massive particles,''
JHEP \textbf{12}, 094 (2015)
doi:10.1007/JHEP12(2015)094
[arXiv:1509.01406 [hep-th]].

\bibitem{Campiglia:2017mua}
M.~Campiglia and R.~Eyheralde,
``Asymptotic $U(1)$ charges at spatial infinity,''
JHEP \textbf{11}, 168 (2017)
doi:10.1007/JHEP11(2017)168
[arXiv:1703.07884 [hep-th]].

\bibitem{Distler:2018rwu}
J.~Distler, R.~Flauger and B.~Horn,
``Double-soft graviton amplitudes and the extended BMS charge algebra,''
JHEP \textbf{08}, 021 (2019)
doi:10.1007/JHEP08(2019)021
[arXiv:1808.09965 [hep-th]].

\bibitem{Pasterski:2016qvg}
S.~Pasterski, S.~H.~Shao and A.~Strominger,
``Flat Space Amplitudes and Conformal Symmetry of the Celestial Sphere,''
Phys. Rev. D \textbf{96}, no.6, 065026 (2017)
doi:10.1103/PhysRevD.96.065026
[arXiv:1701.00049 [hep-th]].

\bibitem{Pasterski:2017kqt}
S.~Pasterski and S.~H.~Shao,
``Conformal basis for flat space amplitudes,''
Phys. Rev. D \textbf{96}, no.6, 065022 (2017)
doi:10.1103/PhysRevD.96.065022
[arXiv:1705.01027 [hep-th]].

\bibitem{Donnay:2018neh}
L.~Donnay, A.~Puhm and A.~Strominger,
``Conformally Soft Photons and Gravitons,''
JHEP \textbf{01}, 184 (2019)
doi:10.1007/JHEP01(2019)184
[arXiv:1810.05219 [hep-th]].

\bibitem{Donnay:2020guq}
L.~Donnay, S.~Pasterski and A.~Puhm,
``Asymptotic Symmetries and Celestial CFT,''
JHEP \textbf{09}, 176 (2020)
doi:10.1007/JHEP09(2020)176
[arXiv:2005.08990 [hep-th]].

\bibitem{Pasterski:2021fjn}
S.~Pasterski, A.~Puhm and E.~Trevisani,
``Celestial diamonds: conformal multiplets in celestial CFT,''
JHEP \textbf{11}, 072 (2021)
doi:10.1007/JHEP11(2021)072
[arXiv:2105.03516 [hep-th]].

\bibitem{Donnay:2022sdg}
L.~Donnay, S.~Pasterski and A.~Puhm,
``Goldilocks Modes and the Three Scattering Bases,''
[arXiv:2202.11127 [hep-th]].

\bibitem{Hinterbichler:2013dpa}
K.~Hinterbichler, L.~Hui and J.~Khoury,
``An Infinite Set of Ward Identities for Adiabatic Modes in Cosmology,''
JCAP \textbf{01}, 039 (2014)
doi:10.1088/1475-7516/2014/01/039
[arXiv:1304.5527 [hep-th]].

\bibitem{Horn:2014rta}
B.~Horn, L.~Hui and X.~Xiao,
``Soft-Pion Theorems for Large Scale Structure,''
JCAP \textbf{09}, 044 (2014)
doi:10.1088/1475-7516/2014/09/044
[arXiv:1406.0842 [hep-th]].

\bibitem{Horn:2015dra}
B.~Horn, L.~Hui and X.~Xiao,
``Lagrangian space consistency relation for large scale structure,''
JCAP \textbf{09}, 068 (2015)
doi:10.1088/1475-7516/2015/09/068
[arXiv:1502.06980 [hep-th]].

\bibitem{Campiglia:2016jdj}
M.~Campiglia and A.~Laddha,
``Sub-subleading soft gravitons: New symmetries of quantum gravity?,''
Phys. Lett. B \textbf{764}, 218-221 (2017)
doi:10.1016/j.physletb.2016.11.046
[arXiv:1605.09094 [gr-qc]].

\bibitem{Campiglia:2016efb}
M.~Campiglia and A.~Laddha,
``Sub-subleading soft gravitons and large diffeomorphisms,''
JHEP \textbf{01}, 036 (2017)
doi:10.1007/JHEP01(2017)036
[arXiv:1608.00685 [gr-qc]].

\bibitem{Campiglia:2014yka}
M.~Campiglia and A.~Laddha,
``Asymptotic symmetries and subleading soft graviton theorem,''
Phys. Rev. D \textbf{90}, no.12, 124028 (2014)
doi:10.1103/PhysRevD.90.124028
[arXiv:1408.2228 [hep-th]].

\bibitem{Conde:2016rom}
E.~Conde and P.~Mao,
``BMS Supertranslations and Not So Soft Gravitons,''
JHEP \textbf{05}, 060 (2017)
doi:10.1007/JHEP05(2017)060
[arXiv:1612.08294 [hep-th]].

\bibitem{Freidel:2021dfs}
L.~Freidel, D.~Pranzetti and A.~M.~Raclariu,
``Sub-subleading Soft Graviton Theorem from Asymptotic Einstein's Equations,''
[arXiv:2111.15607 [hep-th]].

\bibitem{Bondi:1962px} 
  H.~Bondi, M.~G.~J.~van der Burg and A.~W.~K.~Metzner,
  ``Gravitational waves in general relativity. 7. Waves from axisymmetric isolated systems,''
  Proc.\ Roy.\ Soc.\ Lond.\ A {\bf 269}, 21 (1962).
  
\bibitem{Sachs:1962wk} 
  R.~K.~Sachs,
  ``Gravitational waves in general relativity. 8. Waves in asymptotically flat space-times,''
  Proc.\ Roy.\ Soc.\ Lond.\ A {\bf 270}, 103 (1962).

\bibitem{Sachs:1962zza} 
  R.~Sachs,
  ``Asymptotic symmetries in gravitational theory,''
  Phys.\ Rev.\  {\bf 128}, 2851 (1962).

\bibitem{Katz:1985}
	J.~Katz, 
	``A note on Komar's anomalous factor,''
	Class. Quant. Grav. \textbf{2}, 43 (1985).
	
\bibitem{Compere:2019gft}
G.~Comp\`ere, R.~Oliveri and A.~Seraj,
``The Poincar\'e and BMS flux-balance laws with application to binary systems,''
JHEP \textbf{10}, 116 (2020)
doi:10.1007/JHEP10(2020)116
[arXiv:1912.03164 [gr-qc]].
	
\bibitem{Barnich:2011mi}
G.~Barnich and C.~Troessaert,
``BMS charge algebra,''
JHEP \textbf{12}, 105 (2011)
doi:10.1007/JHEP12(2011)105
[arXiv:1106.0213 [hep-th]].
	

\bibitem{Donnay:2021wrk}
L.~Donnay and R.~Ruzziconi,
``BMS flux algebra in celestial holography,''
JHEP \textbf{11}, 040 (2021)
doi:10.1007/JHEP11(2021)040
[arXiv:2108.11969 [hep-th]].

\bibitem{Penna:2015gza}
R.~F.~Penna,
``BMS invariance and the membrane paradigm,''
JHEP \textbf{03}, 023 (2016)
doi:10.1007/JHEP03(2016)023
[arXiv:1508.06577 [hep-th]].

\bibitem{Donnay:2016ejv}
L.~Donnay, G.~Giribet, H.~A.~Gonz\'alez and M.~Pino,
``Extended Symmetries at the Black Hole Horizon,''
JHEP \textbf{09}, 100 (2016)
doi:10.1007/JHEP09(2016)100
[arXiv:1607.05703 [hep-th]].



\end{thebibliography}
\bibliographystyle{JHEP}
\renewcommand{\refname}{Bibliography}
\addcontentsline{toc}{section}{Bibliography}
\providecommand{\href}[2]{#2}\begingroup\raggedright

\end{document}